# Effects of P, As, and Sb heavy doping on band gap narrowing of germanium as light-emitting materials

Zhong-Hua Dai, Yao-Ping Xie*, Yi-Chen Qian, Li-Juan Hu, Xiao-Di Li and Hai-Tao Ma

Key Laboratory for Microstructures and Institute of Materials Science, School of Materials Science and Engineering, Shanghai University, Shanghai 200072, China

Abstract: The n-type tensile-strained Ge can be used as high-efficient light-emitting materials. To reveal the influence of n-type doping on the electronic structure of Ge, we have computed the electronic structure of P, As and Sb doped Ge using first-principles calculation and band unfolding technique. We find that these n-type doping can induce both indirect and direct band gap narrowing, which well reproduce experimental observation that red-shifts occur in photoluminescence spectra of Ge with n-type doping. We reveal that the indirect band gap narrowing is mainly caused by impurity state, while the direct band gap narrowing is a result of lattice distortion induced by the dopant atom. Moreover, we find that it can use $E_g^\Gamma$-$E_g^L$ to explain the voltage increase was needed to reach the same current densities of light emission through the different samples with increasing doping concentrations.

## Introduction

Ge is a semiconductor material` with advantageous properties such as high mobility of holes and electrons, and as one of first generation semiconductor materials Ge has been studied extensively. Moreover, Ge is compatible with existing Si technology [1,2]. Recently, it is demonstrated that strained Ge with n-type doping can be used as a light-emitting material [3-7]. Modulation of the band gap of Ge toward realization of desirable waveband for given applications and promotion of the efficiency of light-emission is now central issues in the field of Si-compatible optoelectronics.

It is known that Ge has an indirect band gap, and there are two closely energy minima in conduction band, i.e. an indirect L valley and a direct Γ valley. The energy of Γ valley is about 140 meV higher than that of L valley at room temperature [2]. The light emitting via indirect transition involving L valley electrons is a phonon-assisted recombination and its luminous efficiency is limited. Promotion of the direct band-to-band transition is a main method to increase the total luminous efficiency of Ge.

To increase the amount of electrons in Γ valley, Sun et al. combine the strain and doping effects to compensate the energy difference between Γ valley and L valley [3-5,8]. Since the indirect-to-direct bandgap transition occurs at 2 % in-plane tensile strain along (001) plane, and at the transition point the bandgap of Ge shrinks to 0.5 eV which corresponds to 2500 nm that is far from the desirable wavelength in telecommunication, they introduce only 0.2-0.25 % tensile strain by using thermal expansion method to reduce the energy difference between Γ valley and L valley [9-11]. In addition, they use n-type heavy doping to fill the indirect L valley with extrinsic electrons, so more injected electrons can stay in Γ valley, which can induce direct transition of electron and induces a high luminous efficiency. Therefore, it is proved that the Ge layer with n-type doping and tensile strain is an efficient light-emitting material.

To further promote the efficiency of light-emission for Ge, profound understandings of the strain and doping on the band structure are necessary. To explore the condition of the indirect to direct band gap transition in Ge, there are many systematical investigations using first-principles methods on the effect of strain on the indirect-to-direct bandgap transition and band gap narrowing (BGN) [12-18]. These studies illustrate the method how to obtain direct band-gap material by exerting strain on Ge. It is found that the (001) biaxial tension is the most efficient among all biaxial approaches to transform Ge into a direct band gap material, the transition point is at 2.91%, and the direct gap in this transition point is 0.33 eV [14]. The [111]-tension is the best choice among all uniaxial approaches for an indirect-to-direct band gap transition of Ge, the transition point is at 5.69 %, and the direct gap in this transition point is also smaller than 0.4 eV [14]. The DFT calculations indicate the Ge can be transferred into a direct band gap material, but its direct band gap is always smaller than 0.4 eV.

In other hand, the heavy doping effect on the direct band emission also has been investigated extensively. E. Kasper et al. [19] reported that Sb doped Ge has photoluminescence peak at 0.73 eV, which is lower than electroluminescence peak, 0.80 eV. It is because electroluminescence stems from the inner intrinsic layers far from the surface, whereas photoluminescence originates from the highly n-doped outmost layers. The band gap topmost layers of Ge are significantly influenced by the Sb doping, and the doping induces BGN effects. Take the doping element P as an example, Camacho-Aguilera et al. [20] proposed that the BGN can be used as a powerful nondestructive method for determining the total active dopant concentration in Ge. Oehme et al. [21] also showed that the BGN increases with the concentration of Sb doping in mild stretched Ge. Schmid et al. [22] systematically investigated Ge with mild tensile strain ranging from 0 % to 0.24 % and with n-type doping ranging from $5 \times 10^{17}$ cm$^{-3}$ to $1 \times 10^{20}$ cm$^{-3}$, and they found there is a distinct band gap narrowing effect with a maximal energy shift of 38 meV for a doping concentration of $1\times10^{20}$ cm$^{-3}$. Schwartz et al. [23] have investigated that electroluminescence of Sb doped Ge, and found the direct and indirect BGN are different.

Since there are many experimental studies focusing on luminous properties of Ge with n-type heavy doping, the underlying mechanism of typical n-type dopant, i.e., P, As, Sb on the BGN is still unclear. It is fundamental to understand these experiments observation for the light-emitting behavior of heavily doped Ge. We noticed that, in all of above experimental studies, the concentration of doping in Ge is nominal concentration. For heavy doping, most of implantation techniques that used to increase the dopant concentration unavoidably introduce other defects, and therefore the total active dopant concentration is not equal to total implanted concentration. The role of active dopant of given elements is difficultly measured from experiments. Therefore, it is particularly necessary to use DFT to explore the efficiency of different types of n-type active dopant. Therefore, we use first-principles calculation and band unfolding technique [24,25] to investigate three typical n-type dopants on the BGN of Ge with and without strain.

**Method**

The electronic structure calculations are undertaken using the Vienna ab initio simulation package (VASP) code based on density functional theory (DFT) [26,27]. The electron exchange and correlation is described within the generalized gradient approximation as parameterized by Perdew, Burke, and Ernzerhof (GGA-PBE) [28], and the interaction between ions and electrons is described using the projector augmented wave method (PAW) [29]. The energy cutoff is chosen as 400 eV for the plane wave expansion of the wave functions.

Here, to save the computational expenditure, we use a 2×2×2 supercell, in which only one atom is substituted by the dopant atom. The concentration of dopant atom is $7.1\times10^{20}$ cm$^{-3}$, which has same maximum order of magnitude with the concentration of heavily doped Ge in experiments. For the 2×2×2 supercell, we use a 5×5×5 k-point grid generated according to the Monkhorst-Pack scheme to sample the Brillouin zone [30]. The GGA+U approach is applied to correct the underestimation of the band gap. With on-site Coulomb parameter U (U=0 eV) and the on-site exchange parameter J (J=3.33 eV), we get a bandgap of 0.745 eV which is in close agreement with the experimental value, 0.74 eV, at 0 K.

The introducing of a supercell will result in a folded band structure due to the shrinkage of the Brillouin zone, which brings in an inconvenience to distinguish the impurity states from bulk states by comparing the difference of band structure from systems with and without doping. Therefore, we use a band unfolding technique named BandUP to unfold the band structure of supercell of Ge. In this method, the spectral function $A(\vec{k};\varepsilon) \equiv \sum_m P_{m\vec{K}}(\vec{k})\delta(\varepsilon - \varepsilon_m(\vec{K}))$ is created to unfold the band structure. In which $P_{m\vec{K}}(\vec{k}) \equiv \sum_n |\langle \psi_{m\vec{K}}^{SC} | \psi_{n\vec{k}}^{PC} \rangle|^2$ denotes the spectral weight, $|\psi_{n\vec{k}}^{PC}\rangle$ and $|\psi_{m\vec{K}}^{SC}\rangle$ are the eigenvectors in the primitive cell and the supercell, respectively. The spectral weight can be obtained by projecting $|\psi_{m\vec{K}}^{SC}\rangle$ on all primitive cell eigenstates $|\psi_{n\vec{k}_i}^{PC}\rangle$ of a fixed $k_i$. Finally, the obtained $A(\vec{k};\varepsilon)$ is regarded as an effective primitive cell projection of the supercell band structure.

We first calculate the band structure of 2×2×2 supercell of pure Ge as is shown in Fig. 1a. Then, we use BandUP to obtain its unfolded band structure which is presented in Fig. 1b. The shape of unfolded band structure shown in Fig. 1b is very different from the folded band structure of 2×2×2 supercell, but fit well with the primitive cell band structure [31]. One can see that it is easy to recognize the indirect and direct band gap (0.745 eV and 0.929 eV) in unfolded band structure but difficult to recognize them from the folded one.

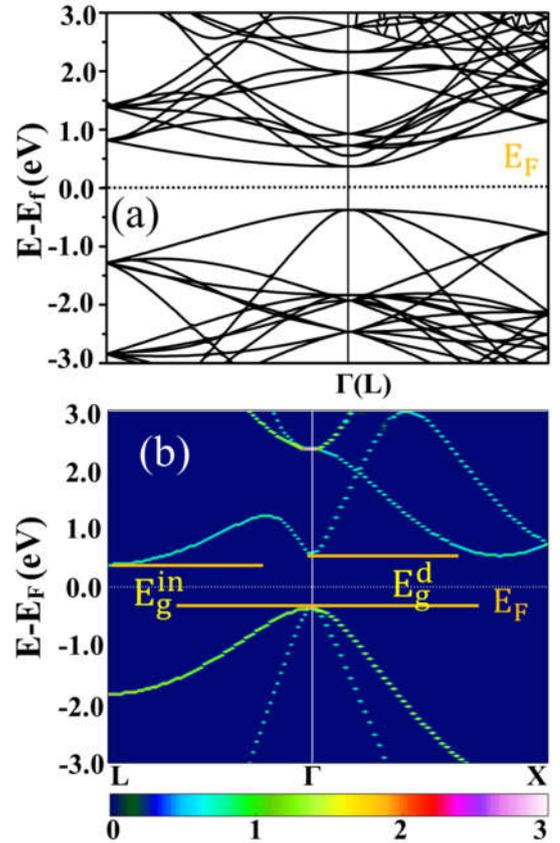

**Figure 1** Band structure of the 2×2×2 supercell of pure Ge: **a** folded band structure; **b** unfolded band structure. The color scale in **b** represents the number of the primitive cell bands crossing the energy interval at a given primitive wave vector.

The direct band gap ($E_g^d$) of heavily n-doped Ge or mild strained Ge (strain < 0.3 %), which corresponds to the peak around 1600 nm in luminescence spectrum, is the difference of energies of Γ point at valence band maximum (VBM) and conduction band minimum (CBM) (see Fig. 1b). It is computed by following equation:

$$E_g^d = E_{CBM}^\Gamma - E_{VBM}^\Gamma$$

where $E_{CBM}^\Gamma$ and $E_{VBM}^\Gamma$ are the energies of electronic state of Γ point at CBM and VBM. The indirect band gap

($E_g^{in}$) of heavily n-doped Ge or mild strained Ge (strain < 0.3 %), which corresponds to the peak around 1800 nm in luminescence spectrum, is the difference of energies of L point at CBM and Γ point at VBM (see Fig. 1b). It is computed by following equation:

$$E_g^{in} = E_{CBM}^L - E_{VBM}^\Gamma$$

where $E_{CBM}^L$ is the energy of electronic state at L point of CBM. The values of BGN of direct band gap in n-doped Ge and strained n-doped Ge ($E_{g\_d}^{dBGN}$, $E_{g\_ds}^{dBGN}$), and values of BGN of indirect band gap in n-doped Ge and strained n-doped Ge ($E_{g\_d}^{iBGN}$, $E_{g\_ds}^{iBGN}$) are defined as following:

$$E_{g\_d}^{dBGN} = E_{g\_d}^d - E_{g0}^d$$

$$E_{g\_ds}^{dBGN} = E_{g\_ds}^d - E_{g0}^d$$

$$E_{g\_d}^{iBGN} = E_{g\_d}^{in} - E_{g0}^{in}$$

$$E_{g\_ds}^{iBGN} = E_{g\_ds}^{in} - E_{g0}^{in}$$

where $E_{g\_d}^d$ and $E_{g\_ds}^d$ are the direct band gap of n-doped Ge and strained n-doped Ge respectively, $E_{g\_d}^{in}$ and $E_{g\_ds}^{in}$ are the indirect band gap of n-doped Ge and strained n-doped Ge respectively, and $E_{g0}^d$ and $E_{g0}^{in}$ are the direct band gap and indirect band gap of Ge in its equilibrium state, respectively.

## Results and discussion

**Table 1** Calculated band gaps of pure Ge and doped Ge. $E_{g\_d}^{\prime dBGN}$ is direct BGN for pure Ge which has the same lattice distortion with the corresponding doped Ge, and $E_g^d$-$E_g^{in}$ denotes the energy difference between Γ and L valley (in meV)

| dopant | $E_g^d$ | $E_g^{in}$ | $E_{g\_d}^{dBGN}$ | $E_{g\_d}^{\prime dBGN}$ | $E_{g\_d}^{iBGN}$ | $E_g^d$-$E_g^{in}$ |
|---|---|---|---|---|---|---|
| Pure Ge | 929 | 745 | - | - | - | 184 |
| P | 919 | 452 | 10 | (9) | 293 | 467 |
| As | 875 | 391 | 54 | (48) | 354 | 484 |
| Sb | 821 | 609 | 108 | (114) | 135 | 211 |

Since the direct band gap is closely related to the main luminous peak around 1600 nm in photon emission spectrum, here we present the values of direct gaps of Ge with doping in Table 1 and give a detail survey of the change of direct band gap induced by doping. All of the direct band gaps of doped Ge are smaller than that of pure Ge, showing a direct BGN effect, and these findings are consistent with the red shift phenomenon produced by direct BGN effect in experiments [4,19-23]. The direct BGNs for P, As and Sb doped Ge are 10 meV, 54 meV and 108 meV, respectively. Based on the calculated band structures in Fig. 2, the direct BGN does not originate from the impurity states. Actually, the direct BGN is induced by the lattice distortion which is introduced by the impurity atom. To illustrate the strain induced by impurity atom, we plot radial distribution function of doped Ge in Fig. 3. From Fig. 3, we find that Sb and As induce much larger lattice distortion than P does. Correspondingly, the BGNs induced by Sb and As are also larger than that induced by P.

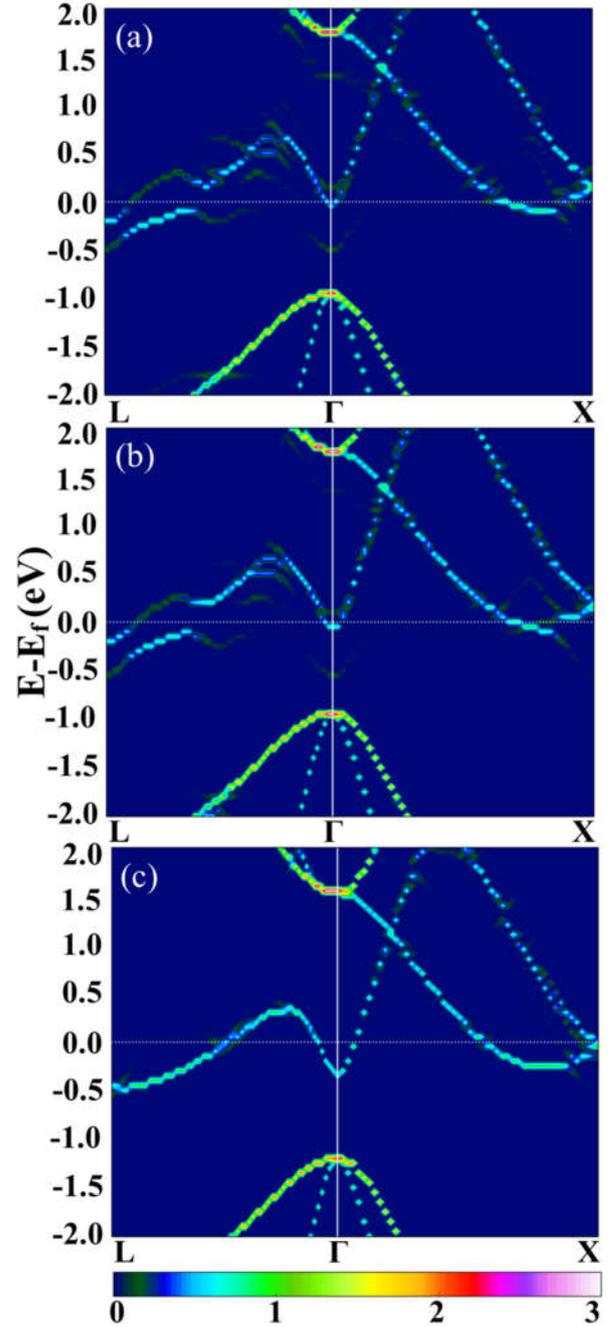

**Figure 2** Band structure of n-type doped Ge: **a** P doped Ge; **b** As doped Ge; **c** Sb doped Ge. The color scale represents the number of the primitive cell bands crossing the energy interval at a given primitive wave vector.

To confirm that the direct BNG derives from the lattice distortion rather than other factors, we compute pure Ge band structures with the same distortion of P, As and Sb doped Ge, respectively. The values of direct BGNs of pure Ge ($E_{g\_d}^{\prime dBGN}$) with lattice distortions as the same extent of inner atomic strain with doped Ge are computed as following equation:

$$E_{g\_d}^{\prime dBGN} = E_{g\_is}^d - E_{g0}^d$$

where $E_{g\_is}^d$ is direct band gap of pure Ge with lattice distortions as the same extent of inner atomic strain with doped Ge, and $E_{g0}^d$ is direct band gap of Ge in its equilibrium state. The values of direct BGNs of pure Ge ($E_{g\_d}^{\prime dBGN}$) with lattice distortions as the same extent of inner atomic strain with doped Ge are listed in bracket within Table 1. It shows that direct BGNs caused by lattice distortion for pure Ge are almost the same with those of corresponding doped Ge. Therefore, we conclude that lattice distortion induced by dopant is the direct reason of the direct BGN.

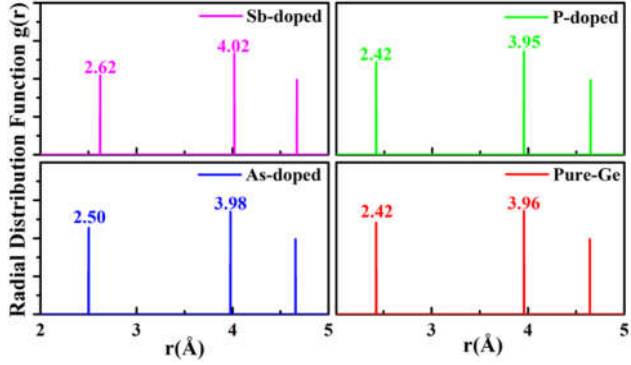

**Figure 3** Comparison of the radial distribution function of Sb, As and P atom in Ge with that of pure Ge.

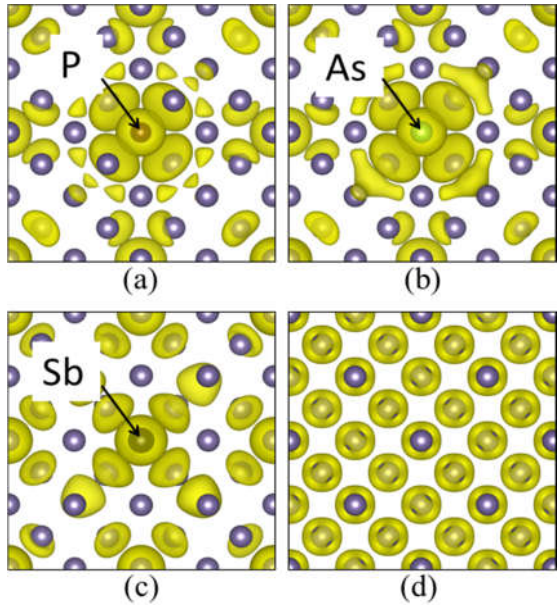

**Figure 4** The partial charge density (PCD) of impurity states at L point for **a** P, **b** As, **c** Sb doped Ge and bulk states at L point for **d** pure Ge.

As shown in Table 1, all values of the indirect band gaps ($E_g^d$) for Ge with these n-type doping are smaller than that of pure Ge, and indicate indirect band gap narrowing also occur. We note that there is few experiments foucus on the light-emitting from indirect band gap due to the difficult of light-emitting via indirect transition of electron. However, the light absorption spectrum shows that indirect BGN induced by the n-type doping does occur. By comparing the band structure of Ge with n-type doping shown in Fig. 2 and band structure of pure Ge shown in Fig. 1, we find that all of the impurity states induced by n-type dopant appear in conduction band. The impurity states mainly locate at L valley, and the spectral weight of impurity states located at Γ valley is very small. Therefore, the impurity state of n-type doping has an obvious influence on the indirect band gap.

In the system with n-type doping, the impurity states and the bulk states of Ge couple with each other, inducing a removal of fourfold degenerate state at L valley. Since P and As dopant have relatively small atomic radius, the valence electrons of P and As have more confinement and the valence electron states are more localized (see Fig. 4a, b). Therefore, the coupling of impurity states and bulk states, the coupling of impurity states from adjacent dopant of P and As dopant are more obvious, inducing that L valley of P and As doped Ge descends more remarkably, as shown in Fig. 2a, b. The shift of L valley of Ge with P and As doping results in BGN of 293 meV and 354 meV for indirect bandgap, respectively. The Sb atom has a large atomic radius, and its valence electron states are delocalized, whose charge density distribution is very similar to that of bulk states (see Fig. 4c, d). Therefore, the coupling of Sb impurity states and bulk states is weak and the shift of L valley is relatively small shown in Fig. 2c. The shift of L valley induced by Sb impurity state only results in an indirect BGN of 135 meV.

Based on all of the calculated band gaps, we find that P and As induce larger indirect BGN than Sb does, indicating that doping of P and As can efficiently change the wavelength of indirect band-to-band transition toward infrared direction than Sb does. In contrast, Sb induces larger direct BGN than P and As do, indicating that doping Sb can efficiently change the wavelength of direct band-to-band transition toward infrared direction than P and As do.

The energy difference between L valley and Γ valley ($E_g^\Gamma - E_g^L$) is also an important parameter. We find that the values of $E_g^\Gamma - E_g^L$ of Ge with dopping are all larger than that of pure Ge. It can be used to explain that doping lead to higher voltages needed for excitation of electrons in electroluminescence [23]. Schwartz showed that the voltage increase was needed to reach the same current densities of light emission through the different samples with increasing doping concentrations [23]. This increasing voltage cannot be explained by the increase of extrinsic electrons, and it only can be explained by the increase of ($E_g^\Gamma - E_g^L$) by Sb doping. From the values of $E_g^\Gamma - E_g^L$ in Table 1, we also can infer that not only Sb doping, but also P, As will increase the voltages needed for excitation of electrons in electroluminescence.

Further, one can infer that the increase of $E_g^\Gamma - E_g^L$ makes Ge need more extrinsic electrons to fill in L valley and reduces the amount of electrons pumped into Γ valley, and therefore it reduces the direct band-to-band transtion and the luminous efficiency. Thus, the larger the value of $E_g^\Gamma - E_g^L$ is, the lower luminous efficiency is. The value of $E_g^\Gamma$-$E_g^L$ for Sb doped Ge, 211 meV, which is only slightly larger than that of pure Ge, is much smaller than those of P and As doped Ge. Therefore, we can predict the Sb doped Ge has a higher luminous efficiency than P and As doped Ge with the same active purity concentration of doping.

For the application of Ge as light-emitting materials realizing telecommunication wavelength of 1550 nm used in the third optical communication window, the n-type doped Ge is usually under the strain about 0.3 %. Therefore, to reveal the combined effect of doping and tensile strain on the band gap of Ge, we compute doped Ge with 0.3% (001) biaxial tensile strain. Table 2 shows that the strain on doped Ge further increases both indirect and direct BGN. The values of direct and indirect band gap ($E_g^d$ and $E_g^{in}$ in Table 2) of Ge with both strain and doping are all smaller than these of Ge only with doping. Naturally, the BGN ($E_{g\_ds}^{dBGN}$ and $E_{g\_ds}^{iBGN}$) of Ge with both strain and doping are smaller than the BGN of Ge only with doping. To reveal the effect of strain on Ge with different dopant, we computed the values of BGN ($E_{g\_s\prime}^{dBGN}$ and $E_{g\_s\prime}^{inBGN}$) only induced by strain in doped Ge according to following equation:

$$E_{g\_s\prime}^{dBGN} = E_{g\_ds}^d - E_{g\_d}^d$$

$$E_{g\_s\prime}^{iBGN} = E_{g\_ds}^{in} - E_{g\_d}^{in}$$

where $E_{g\_ds}^d$ and $E_{g\_ds}^{in}$ are direct and indirect band gaps of Ge with doping and strain respectively, while $E_{g\_d}^d$ and $E_{g\_d}^{in}$ are direct and indirect band gaps of Ge just with doping respectively.

**Table 2** Calculated band gaps and BGN of 0.3%(001) biaxial tensile strained Ge with and without doping. $E_{g\_s\prime}^{dBGN}$ and $E_{g\_s\prime}^{iBGN}$ are strain induced direct and indirect BGN in doped Ge (in meV)

| doping | $E_g^d$ | $E_g^{in}$ | $E_{g\_ds}^{dBGN}$ | $E_{g\_s\prime}^{dBGN}$ | $E_{g\_ds}^{iBGN}$ | $E_{g\_s\prime}^{iBGN}$ |
|---|---|---|---|---|---|---|
| Pure Ge | 863 | 704 | | 66 | | 41 |
| P | 862 | 411 | 67 | 57 | 335 | 42 |
| As | 818 | 351 | 111 | 57 | 394 | 40 |
| Sb | 759 | 570 | 170 | 62 | 175 | 40 |

We find that the strain induced indirect BGNs of Ge with different dopant are almost the same, indicating the effect of strain on the BGNS is independent on the dopant. The reason is that the indirect BGN mainly contributed by the impurity states. The influence of strain on impurity state should be insensitive to the type of dopant, and therefore the BGN is not influenced by the type of dopant. In contrast, the strain induced direct BGNs are different, indicating the effect of strain on the direct BGNs is dependent on the dopant. The reason is that direct BGN mainly induced by the inner strain of doped Ge, and the strain can also change the inner strain around the dopant atom. The strain can give more inner strain variation on the lager dopant atoms in Ge, such as Sb, and give less strain variation on smaller dopant atom in Ge such as P and As, and therefore, the values of strain induced direct BGN of Ge with Sb doping are larger than that of Ge with P and As.

Though the influence of strain on direct and indirect band gap are different, the strain induced indirect BGNs, about 40 meV, are always smaller than strain induced direct BGNs, among 57~62 meV. As a result, the strain decreases of the values of $E_g^\Gamma$-$E_g^L$, indicating that the strain can decrease the efficiency to fill in L valley with extrinsic electrons and increase the electrons in Γ valley and promote luminous efficiency of Ge both with and without doping. Therefore, we here can illustrate the role of strain used in the n-doped Ge for light-emitting materials is to decrease $E_g^\Gamma$-$E_g^L$, which is same with previous studies of the effect of stain on pure Ge [12-16].

To discuss the effect of these three n-type dopants on the luminescence efficiency, many factors should be considered. First, the ability of providing electrons of dopant should be considered. Usually, the electronegativity reflects the ability of an atom attracting an electron to itself. Therefore, the smaller the value of dopant electronegativity is, the stronger the ability of dopant providing electrons is. Second, the change of value of $E_g^\Gamma$-$E_g^L$ by dopant that mentioned previous section should be considered. The smaller the value of $E_g^\Gamma$-$E_g^L$ is, the more the amount of electrons can be pumped in the Γ valley from L valley. A smaller value of $E_g^\Gamma$-$E_g^L$ also needs a lower voltages for excitation of electrons in electroluminescence. Third, the order of difficulty of implantation of dopant should also been considered. The difficulty of implantation of dopant can be reflected by cohesive energy of dopant. The cohesive energy of dopant atom in Ge is the bonding energy between dopant atom and Ge atom, and it can be calculated from following equation:

$$En_{coh} = En_{dcoh} - En_{Gecoh}.$$

where $En_{dcoh}$ is the total energy of dopant system and $En_{Gecoh}$ is the total energy of pure Ge. Now, we list the electronegativity, the values of $E_g^\Gamma$-$E_g^L$ of Ge with 0.3 % strain and doping, and cohesive energy of per dopant atom in Ge in Table 3.

The electronegativity of Sb is the smallest among these three dopants, indicating that Sb have the strongest ability to provide electrons for luminescence. The $E_g^\Gamma$-$E_g^L$ of Sb is the smallest among these three dopants, indicating amount of electrons needed to cram the L valley for Sb doped Ge is smaller than that needed for P and As doped Ge. Moreover, it is easier to inject electrons into the Γ valley from L valley of Sb doped Ge than P and As doped Ge; because the voltages needed for excitation of electrons in electroluminescence for Sb doped Ge are smaller than P and As doped Ge. The cohesive energy of P, As, Sb in Ge are all positive, indicating that the implantation of these atoms into Ge are endothermic processes. The cohesive energy of Sb is the largest, indicating the implantation of Sb in Ge is most difficult.

Therefore, we can conclude that though the implantation of Ge is most difficult among these three n-type dopants, the effect of active Sb dopant has the highest ability to enhance luminescence efficiency of n-type doping strained Ge compared with P As doping at the same active doping concentration.

Table 3 The electronegativity (n) from ref [32], and calculated $E_g^\Gamma$-$E_g^L$, $En_{coh}$ of P, As, and Sb in Ge

| dopant | n | $E_g^\Gamma$-$E_g^L$ | $En_{coh}$ |
|---|---|---|---|
| P | 2.19 | 451 meV | 1.74 eV |
| As | 2.18 | 466 meV | 2.49 eV |
| Sb | 2.05 | 189 meV | 3.45 eV |

## Conclusion

In summary, we investigate the electronic structure of P, As and Sb doped Ge using the first-principles method and band unfolding technique. It is found that all of these doping induce both indirect and direct BGN effects which are consistent with experiment observation. Furthermore, we reveal the influence mechanism of these n-type doping on Ge band structure. We find the indirect BGN originates from the impurity state of n-type dopant that appears in L valley, while the direct BGN originates from the lattice distortion induced by the dopant. Specially, we find that it can use $E_g^\Gamma$-$E_g^L$ to explain the voltage increase was needed to reach the same current densities of light emission through the different samples with increasing doping concentrations. By comparing the electronegativity, energy difference between Γ and L valley and cohesive energy, we find that Sb dopant has the highest ability to enhance luminescence efficiency of n-type doping strained Ge compared with P As doping at the same active doping concentration.

**Acknowledgements**

This work is supported by Science and Technology Commission of Shanghai Municipality (15ZR1416000), China Academy of Engineering Physics Joint Funds of National Natural Science Foundation of China (U1530115) and National Science Foundation of China (51301102). High performance computing resources are provided by the Ziqiang Supercomputer Center at Shanghai University. Y. P. thanks for the critical discussion with Doctor Li Yong-Hua and Yin Wan-Jian.

* Corresponding author. Email address: ypxie@shu.edu.cn